\documentclass[10pt,letterpaper,twocolumn]{article} 

\usepackage{ol2}
\usepackage[draft]{hyperref}
\usepackage{amsmath}

\begin{document}

\twocolumn[ 

\title{Zak phase of photons in optical waveguide lattices}


\author{Stefano Longhi}

\address{Dipartimento di Fisica, Politecnico di Milano and Istituto di Fotonica e Nanotecnologie del Consiglio Nazionale delle Ricerche, Piazza L. da Vinci 32, I-20133 Milano, Italy (stefano.longhi@polimi.it)}

\begin{abstract}
Zak phase, i.e.  the Berry phase acquired during an adiabatic motion of a Bloch particle across the Brillouin zone, provides a measure of the topological invariant  of Bloch bands in one-dimensional crystalline potentials. Here a photonic structure, based  on engineered lattices of evanescently-coupled optical waveguides, is proposed to detect Zak phase difference of photons undergoing Bloch oscillations in topologically distinct Bloch bands of dimerized superlattices. 
\end{abstract}

\ocis{350.1370, 230.7370, 000.1600, 050.5298}


 ] 

\noindent

Berry phase plays an important role in a variety
of physical fields, including molecular and condensed-matter physics \cite{Berr1,Berr2}, and optics \cite{Berr3}.  An important case is the Berry phase in Bloch bands of a crystalline solid. When the quasi-momentum of a Bloch particle is forced to span a closed path in momentum space by application of a magnetic or an electric field, the Bloch state acquires a Berry phase, as recognized by  Zak in a seminal paper \cite{ Zak}. Zak phase is useful in the understanding of a host of solid state
phenomena, such as  the
problem of electric polarization in dielectrics \cite{uff1}, the theory
of the integer quantum Hall effect  \cite{uff2}, and the existence of protected edge states \cite{P1,P2}. It also plays an important role in the formation of
Wannier-Stark ladders \cite{Zak,Berr2}.  While Berry phase generally requires a multidimensional parameter space, Zak phase can appear even in a one-dimensional (1D) crystal. Such a remarkable property follows from the torus topology of the Brillouin zone. A paradigmatic 1D model to study Zak phase is provided by the the Su-Schrieffer-Heeger (SSH) model of polyacetylene \cite{poly}, which exhibits two topologically distinct phases \cite{P2,cazz1,cazz2}. The distinct topological character of the two phases is reflected in distinct Zak phases and in the existence of edge states in one of the two phases \cite{P2}. While direct imaging of edge states and the topological properties of one- and two-dimensional crystals and quasi-crystals have been proposed and  demonstrated in a few recent experiments, using either photons \cite{figa1,figa2,figa3} or cold atom systems \cite{Lewen},  the possibility to directly detect the Berry phase in an  interference experiment, as originally suggested by Zak \cite{Zak}, has been so far limited to ultracold atoms in optical lattices \cite{BlochPRL,unpub}. \par In this Letter we suggest a simple optical system, based on lattices of circularly-curved optical waveguides, where the Zak phase  can be detected interferometrically for photons undergoing Bloch oscillations (BOs) \cite{Pesch,BO1,BO2,BO3,BO4} in topologically-distinct Bloch bands.\\
\begin{figure}[htb]
\centerline{\includegraphics[width=8.2cm]{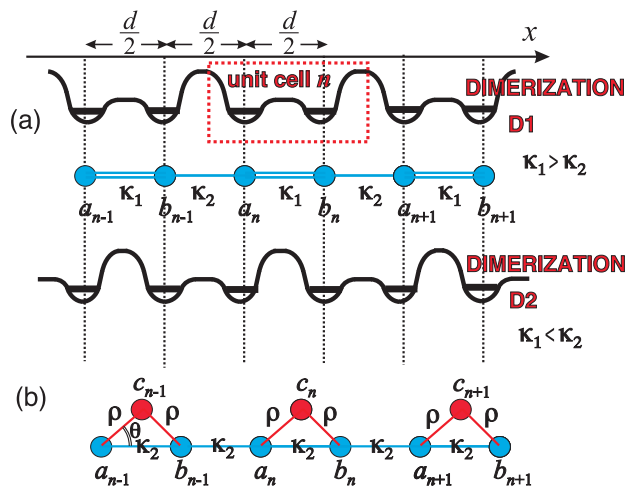}} \caption{
(Color online)  (a) Schematic of a one-dimensional superlattice with period $d$ that realizes the SSH model (3) of polyacetylene. The lattice unit cell contains two sites, $a_n$ and $b_n$. The two realizations D1 and D2, with either $\kappa_1 > \kappa_2$ or $\kappa_1 < \kappa_2$, correspond to two distinct topological phases. The Zak phases in the topological states D1 and D2 differ by $\pi$. (b) Photonic realization of the SHH model based on an array of coupled optical waveguides with equal spacing $d/2$. The auxiliary waveguides at sites $c_n$ are used to control the hopping rates.}
\end{figure}
Let us consider light propagation at wavelength $\lambda$ in a 1D photonic lattice with lattice period $d$, substrate refractive index $n_s$, and periodic index change $\Delta n(x)=\Delta n(x+d)$ in the transverse $x$ direction. In the presence of  a transverse refractive index gradient $Fx$, in the scalar and paraxial approximations light propagation along the paraxial direction $z$ of the lattice is described by the Schr\"{o}dinger-like equation for the field envelope $\psi(x,z)$ \cite{BO4,LonghiPRB}
\begin{equation}
i \hbar \partial_z \psi(x,z)=\hat{H}_0 \psi(x,z)-Fx \psi(x,z)
\end{equation} 
where $\hat{H}_0=-(\hbar^2/2n_s) \partial^2_x+V(x)$, $V(x)=-\Delta n(x)$ is the periodic lattice potential, and $\hbar= \lambda / (2 \pi)$ is the reduced wavelength. Owing to the transverse index gradient, an optical beam injected into the lattice undergoes BOs and Zener tunneling \cite{BO1,BO2,BO3,BO4,Felix}. The  spatial period of BOs is given by $z_B= 2 \pi \hbar /dF=\lambda/dF$. If the gradient is weak and Zener tunneling (ZT) among different Bloch bands is negligible, an initial wave packet $\psi(x,0)$ belonging to the $\alpha$-{\it th} band of the lattice, $\psi(x,0)=\psi_{\alpha}(x,0)$, reproduces itself after one BO cycle, apart from an additional phase $\gamma_{\alpha}=\gamma_{\alpha}^{(D)}+\gamma_{\alpha}^{(Z)}$, i.e. $\psi_{\alpha}(x,z_B)=\psi_{\alpha}(x,0) \exp(i \gamma_\alpha)$ \cite{Zak,LonghiPRB}. The accumulated phase $\gamma_{\alpha}$ comprises the {\it dynamical} phase $\gamma_{\alpha}^{(D)}= -(1/F) \int_{-\pi /d}^{\pi / d} dk E_{\alpha}(k)$, and the {\it geometric} (Zak) phase 
\begin{equation}
\gamma_{\alpha}^{(Z)}=\int_{- \pi/d}^{\pi/d} X_{\alpha, \alpha}(k),
\end{equation}
 where $E_{\alpha}(k)$ is the energy dispersion relation of the $\alpha$-{\it th} lattice band, $X_{\alpha,\alpha}(k)=(2 \pi i /d) \int_{0}^{d} dx u^*_{\alpha}(x,k) \partial_k u_{\alpha}(x,k)$, and $u_{\alpha}(x,k)$ is the periodic part of the Bloch function $\phi_{\alpha}(x,k)=u_{\alpha}(x,k) \exp(ikx)$, with $\hat{H}_0 \phi_{\alpha}(x,k)=E_{\alpha}(k) \phi_{\alpha}(x,k)$.  It should be noted that the value of the Zak phase $\gamma_{\alpha}^{(Z)}$ depends on the origin of the unit cell, and it changes according to $\gamma_{\alpha}^{(Z)} \rightarrow \gamma_{\alpha}^{(Z)}+ 2 \pi (a/d)$ for a translation of the origin by $a$. Such a change is consistent with the fact that, after translation of the entire lattice by $a$ in the gradient potential $Fx$, the Wannier-Stark ladder spectrum is shifted in energy by  $Fa$ and thus an additional phase accumulation, over one BO cycle, by $Fa z_B/ \hbar=2 \pi a/d$ should occur \cite{Berr1}. However, the difference of Zak phases $\gamma_{\alpha}^{(Z)}-\gamma_{\beta}^{(Z)}$ between two different bands $\alpha$ and $\beta$ is an invariant. An important case is the one of a dimerized superlattice with two sites per unit cell \cite{P2,cazz1,cazz2,unpub}, see Fig.1(a).  In the tight-binding and nearest-neighbor approximations, this lattice reproduces the SSH model of polyacetylene \cite{poly} with Hamiltonian 
 \begin{equation}
 \hat{H}_0=-\sum_n  (\kappa_{1} \hat{a}_n^{\dag} \hat{b}_n+\kappa_2 \hat{a}_n^{\dag} \hat{b}_{n-1}+H.c.),
 \end{equation}
 \begin{figure}[htb]
\centerline{\includegraphics[width=8.3cm]{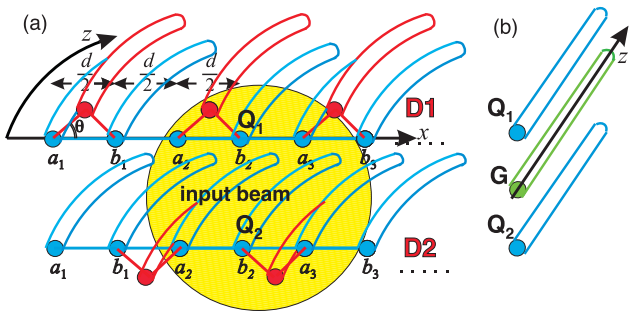}} \caption{
(Color online)  (a) Set-up to detect the $\pi$ phase shift of Zak phases in the topological states D1 and D2 of the SSH model. The two vertically-displaced circularly-bent  waveguide arrays realize the D1 and D2 topological phases. The arrays are illuminated at the input plane by a broad Gaussian beam at normal incidence (the circular spot in the figure). (b) Schematic of the waveguide tritter used to detect the $\pi$ phase shift after one BO cycle.}
\end{figure}

  where $\kappa_{1,2}$ are the hopping rates of alternating lattice sites and $\hat{a}^{\dag}_{n}$ ($\hat{b}^{\dag}_{n}$) are the photon creation operators on the sublattice site $a_n$ ($b_n$) in the $n$-th lattice cell. The SSH model is a two-band model 
with the dispersion relations $E_1(k)=-\sqrt{\kappa_1^2+\kappa_2^2+2 \kappa_1 \kappa_2 \cos (kd)}$ and $E_2(k)=-E_1(k)$ for the two energy bands. For $\kappa_1 \neq \kappa_2$, the two bands are separated by a gap. The Hamiltonian $\hat{H}_0$ is known to exhibit two topologically distinct
phases, referred to as dimerization D1 for $\kappa_1 > \kappa_2$, and dimerization D2 for $\kappa_1< \kappa_2$, see Fig.1(a). For a finite chain of dimers, the topological phase D2 admits of two edge bound states with energy at the midgap, whereas the topological phase D1 does not \cite{P2}. The distinct topological character of the two phases is reflected in different values of their Zak phases \cite{P2}. Assuming the origin of the unit cell as in Ref.\cite{P2}, one has $X_{1,1}(k)=X_{2,2}(k)= -(1/2) (d \theta / dk)$, where $\theta(k)$ is the phase of the complex number $\epsilon(k)=\kappa_1+\kappa_2 \exp(ikd)$.  From Eq.(2), it follows that the Zak phase is equal for the two bands and given by $\gamma_1^{(Z)}=\gamma_2^{(Z)}=0$ for the dimerization D1, and $\gamma_1^{(Z)}=\gamma_2^{(Z)}=\pi$ for the dimerization D2. The observation of the $\pi$ phase shift between the Zak phases in the two topological phases has been recently reported in Ref.\cite{unpub} for cold atoms in optical lattices using a combination of
BOs and Ramsey interferometry. Here we suggest a simple waveguide-based photonic structure where the $\pi$ phase difference of Zak phases can be directly revealed for photons as well. To realize the SSH model of Fig.1(a), we consider a 1D array of equal optical waveguides, equally spaced with spacing $d/2$, with coupling rate $\kappa_2$ between adjacent waveguides. The change of the coupling rate at alternating sites, from $\kappa_2$ to $\kappa_1$, is obtained by adding a set of auxiliary waveguides at sites $c_n$ with a propagation constant detuning $U$ from the waveguides in the main lattice and coupling rate $\rho$, see Fig.1(b). The Hamiltonian of the lattice of Fig.1(b) is given by 
\begin{eqnarray}
\hat{H}_0 & = & -\sum_n \kappa_2(\hat{a}_n^{\dag} \hat{b}_{n}+\hat{a}_n^{\dag} \hat{b}_{n-1}+H.c.) \\
& - &\sum_n \rho (\hat{a}_n^{\dag} \hat{c}_n+\hat{b}_n^{\dag} \hat{c}_n +H.c.)+\sum_n U \hat{c}_n^{\dag} \hat{c}_n. \nonumber
\end{eqnarray}
 In the large detuning limit $|U| \gg \kappa_2, \rho$, the effect of the auxiliary waveguides is to modify the coupling $\kappa_2$ between sites $a_n$ and $b_n$ to the effective value 
 \begin{equation}
 \kappa_1=\kappa_2+\rho^2/U.
 \end{equation}
where the additional contribution $\rho^2/U$ comes from second-order tunneling \cite{Corrielli}. This result can be derived by application of second-order perturbation theory to the Heisenberg equations of motion for the operators $\hat{a}_n$, $\hat{b}_n$ and $\hat{c}_n$, assuming  $\kappa_2/U$ as a small parameter \cite{Corrielli}. Basically, in the large detuning limit $|U| \gg \kappa_2, \rho$ the operators $\hat{c}_n$ can be adiabatically eliminated from the Heisenberg equations, obtaining $\hat{c}_n \simeq(\rho/U)(\hat{a}_n+\hat{b}_n) $. The resulting equations for the operators $\hat{a}_n$ and $\hat{b}_n$ can then be derived from the effective Hamiltonian 
\begin{equation}
\hat{H}_{e}  =  -\sum_n  (\kappa_{1} \hat{a}_n^{\dag} \hat{b}_n+\kappa_2 \hat{a}_n^{\dag} \hat{b}_{n-1}+H.c.) 
-   \frac{\rho^2}{U}\sum_n (\hat{a}^{\dag}_n \hat{a}_n+\hat{b}^{\dag}_n \hat{b}_n) \;\;\;\;
\end{equation}
where $\kappa_1$ is given by Eq.(5).
\begin{figure}[htb]
\centerline{\includegraphics[width=8.5cm]{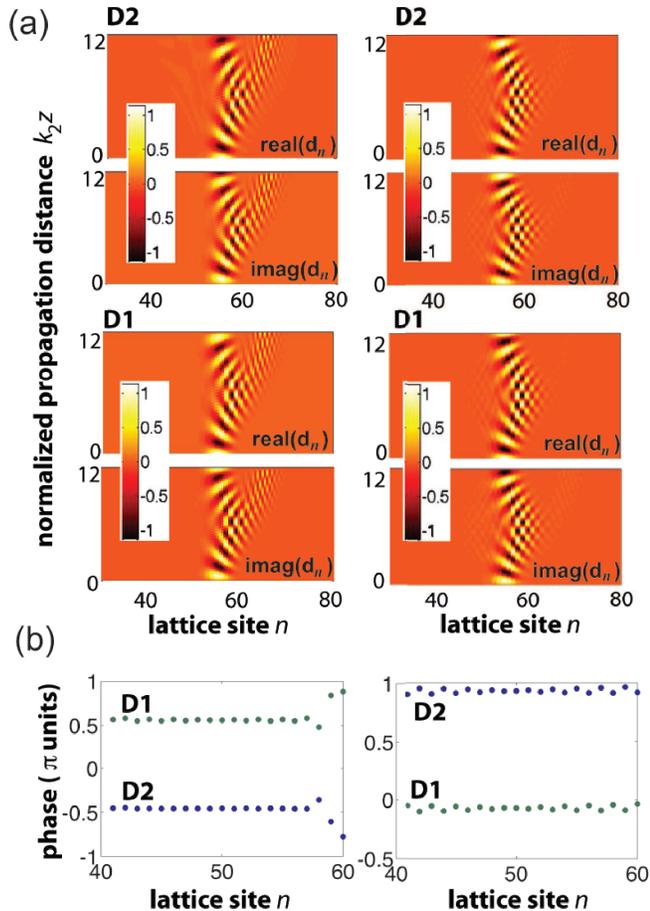}} \caption{
(Color online)  (a) Evolution of the site occupation amplitudes $d_n(z)$ (real and imaginary parts) over one BO cycle [$d_n=a_{(n+1)/2}$ for $n$ odd, $d_n=b_{n/2}$ for $n$ even] in the two lattices D1 and D2 of Fig.2(a) for a Gaussian beam excitation. Parameter values are given in the text. (b) Phase of $d_n$ at $z=z_B$, i.e. after one BO cycle, for the dimerizations D1 and D2. The left panels in the figure refer to the model (4), whereas the right panels are the predictions of the SSH model (3).}
\end{figure}
\begin{figure}[htb]
\centerline{\includegraphics[width=8cm]{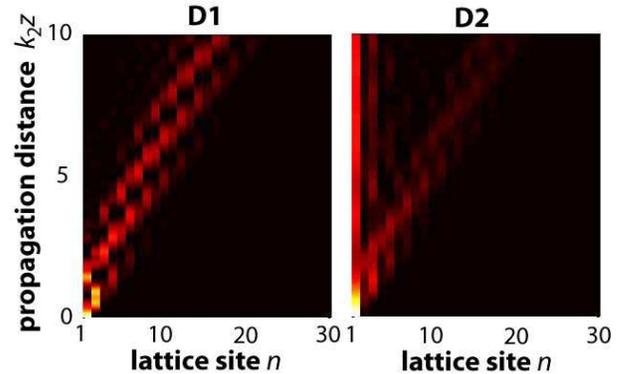}} \caption{
(Color online)  Evolution of the site occupation probabilities $|d_n(z)|^2$ in the straight arrays D1 and D2 corresponding the single-site excitation of the left-edge site $a_1$. The light trapped at the boundary of the array D2 is the clear signature of the existence of an edge state.}
\end{figure}
 $\hat{H}_{e}$ is equivalent to the SSH Hamiltonian (3), apart from the unimportant shift of the energy by $-\rho^2/U$. BOs are obtained by bending the axis of waveguides, the index gradient $F$ being given by $F=n_s/R$, where $R$ is the bending radius of curvature \cite{BO4,Felix}.  To detect the $\pi$ phase difference for the topological phases D1 and D2, we consider two waveguide arrays, in which the auxiliary waveguides are displaced by one unit, see Fig.2(a). We assume $U>0$, so that $\kappa_1> \kappa_2$. The upper and lower arrays in Fig.2(a) realize the phases D1 and D2, respectively. The two arrays are vertically displaced so that evanescent coupling among waveguides belonging to D1 and D2 is negligible over one BO cycle. It should be noted that a simpler realization of the two dimerization phases D1 and D2, obtained without using the auxiliary waveguides and merely alternating the distances $d_1$ and $d_2$ between adjacent guides in the main arrays (with $d_1+d_2=d$), would introduce a shift in the origin of the unit cell in the two states for $d_1 \neq d_2$, and hence an additional phase shift of the field in view of the dependence of the Zak phase on the origin of the unit cell. Conversely, the structure proposed in Fig.2(a) does not shift the position of the sites $a_n$ and $b_n$ in the two lattices, which are equally spaced by $d/2$. The two arrays D1 and D2 are illuminated by a broad Gaussian beam at normal incidence, as shown in Fig.2(a). The beam excites two wave packets in the two arrays belonging to the lowest miniband, which undergo BOs with negligible ZT into the upper miniband for a weak bending. After the propagation distance $z_B=\lambda/(dF)$, self-imaging is observed, however a $ \pi$ phase shift is expected to occur between the wave packets in arrays D1 and D2. In fact, the dynamical phase $\gamma_1^{(D)}$ acquired by the photons is the same in the arrays D1 and D2, whereas their geometric phases differ by $\pi$ because of the different topologies of the lattices. We checked this feature by numerically solving the coupled-mode equations for the full Hamiltonian (4) with an added transverse index gradient, and compared the numerical results with the predictions based on the SSH model [Eq.(3)]. As an example, in Fig.3(a) we show the  evolution of  site occupation amplitudes $d_n$ over one BO cycle in the waveguides belonging to D1 and D2, where $d_n=a_{(n+1)/2}$ for $n$ odd and $d_n=b_{n/2}$ for $n$ even. 
Parameter values used in the simulations are $\rho / \kappa_2=3.4641$, $U/ \kappa_2=8$, and  $Fd/ \hbar=0.5 \kappa_2$,  corresponding to $\kappa_1=2.5 \kappa_2$, which are feasible to be implemented in waveguide lattices manufactured by fs laser writing \cite{BO3,Corrielli,refe}. The lattices comprise $N=50$ unit cells, i.e. $2N=100$ total lattice sites $a_n$, $b_n$. The input excitation is a Gaussian beam with spot size $w=2d$. The appearance of the $\pi$ phase shift between the two wave packets in D1 and D2 is clearly shown in Fig.3(b), which depicts the phases of $d_n$ at $z=z_B$. The shift of the absolute phases in right and left panels of Fig.3(b) is due to the offset $U/\rho^2$ entering in the effective Hamiltonian (6), whereas
the small phase fluctuations after the BO cycle are ascribable to a non-negligible (though small) excitation of the higher lattice miniband by the input beam.  
The larger ZT observed in the left panel of Fig.3(a) is due to the relatively low value of $U/ \kappa_2$ used in the simulations.
ZT is found to decrease as $U/ \kappa_2$ is increased, however it is not detrimental for the observation of the $\pi$ phase shift. We note that larger values of $U/\kappa_2$ could be difficult to be implemented using e.g. fs laser writing, where the range of detuning $U$ attainable is limited by the laser writing speed \cite{Corrielli}.  
The $\pi$ phase shift between the two wave packets in D1 and D2 can be  checked by considering the tritter scheme depicted in Fig.2(b). The waveguides in the structure of Fig.2(a) are terminated at the distance $z_B$ from the input plane,  except for two waveguides $Q_1$ and $Q_2$ in D1 and D2, corresponding to the same lattice site $a_n$ (or $b_n$) in one illuminated unit cell, see Fig.2(b). At $z>z_B$, the waveguides $Q_1$ and $Q_2$ are almost decoupled because 
of their wide separation. Insertion of an auxiliary waveguide G between $Q_1$ and $Q_2$, i.e. realization of a tritter system [see Fig.2(b)], enables evanescent coupling and light transfer among the three waveguides $Q_1$, $G$ and $Q_2$. If the excited modes in waveguides $Q_1$ and $Q_2$ at $z=z_B$  were in phase, the intermediate waveguide $G$ would be illuminated.  Conversely, if the excited modes in waveguides $Q_1$ and $Q_2$ at $z=z_B$ have a $ \pi$ phase shift, the intermediate waveguide $G$ in the tritter is not illuminated, since the initial excitation corresponds to an eigenstate of the tritter with a dark spot in the middle waveguide $G$. Hence the absence of illumination of waveguide $G$ in the tritter is the signature of the $\pi$ shift of Zak phases in the two topological phases D1 and D2. According to Ref.\cite{P2}, the difference of Zak phases in the two topological phases D1 and D2  is closely related to the existence of edge states in one phase (D2), but not in the other one (D1). This is clearly shown in Fig.4, where the  numerically-computed evolution of the site occupation probabilities $|d_n(z)|^2$ is shown corresponding to the initial single-site left-edge excitation of the arrays D1 and D2 of Fig.2(a), in the absence of the external index gradient, i.e. for the straight arrays.\\ Our proposed scheme could be realized using waveguide arrays manufactured in fused silica by fs laser writing \cite{BO3,Corrielli,refe}. For  probing light in the red ($\lambda=633$ nm, $n_s=1.45$) and assuming a lattice period $d=36 \; \mu$m, from the data of Ref.\cite{Corrielli} one can estimate $\kappa_2 \simeq 1.15 \; {\rm cm}^{-1}$. The coupling $\rho=3.4641 \kappa_2 \simeq 4 \; {\rm cm}^{-1}$ of auxiliary waveguides is obtained for an angle $\theta \simeq 0$, wheraes  the propagation constant detuning $U=8 \kappa_2 \simeq 9.2 \; {\rm cm}^{-1}$ can be realized by varying the speed of the fs writing laser beam \cite{Corrielli}. The forcing $Fd/ \hbar=0.5 \kappa_2$, used in the simulations of Fig.3, corresponds to a bending radius $R=n_s/F \simeq 9$ m and to a BO period $z_B \simeq 10.9$ cm.\par
In conclusion, a simple integrated optical structure has been proposed to measure the Zak phase of photons undergoing Bloch oscillations in topologically distinct Bloch bands of the SSH model. It is envisaged that our suggested scheme could be extended to investigate the interplay of Zak phase and topological states in other lattice systems, for example in graphene-like structures \cite{P2} by use of honeycomb photonic lattices \cite{figa3} or in non-Hermitian dimerized lattices \cite{Levi} using lossy guides.

\newpage

\footnotesize {\bf References with full titles}\\
\\
\noindent
1. R. Resta, "Manifestations of Berry's phase in molecules and condensed matter", J. Phys.: Condens. Matter {\bf 12}, R107 (2000).\\
2. D. Xiao,  M.-C. Chang, and Q. Niu, "Berry phase effects on electronic properties", Rev. Mod. Phys. {\bf 82}, 1959 (2010).\\
3. J. C. Guti\'{e}rrez-Vega, "Pancharatnam-Berry phase of optical systems", Opt. Lett. {\bf 36}, 1143 (2010).\\
4. J. Zak, "Berry's Phase for Energy Bands in Solids", Phys. Rev. Lett. {\bf 62}, 2747 (1989).\\
5. R. D. King-Smith and D. Vanderbilt, "Theory of polarization of crystalline solids", Phys. Rev. B {\bf 47}, 1651 (1993).\\
6. J.E. Avron, D. Osadchy, and R. Seiler, "A Topological Look at the Quantum Hall Effect", Phys. Today {\bf 56}, 38 (2003).\\
7. S. Ryu and Y. Hatsugai, "Topological Origin of Zero-Energy Edge States in Particle-Hole Symmetric Systems", Phys. Rev. Lett. {\bf 89}, 077002 (2002).\\
8. P. Delplace, D. Ullmo, and G. Montambaux, "Zak phase and the existence of edge states in graphene", Phys. Rev.
B {\bf 84}, 195452 (2011).\\
9. W. P. Su, J. R. Schrieffer, and A. J. Heeger, "Solitons in Polyacetylene", Phys. Rev. Lett. {\bf 42}, 1698 (1979).\\
10. S. Ryu and Y. Hatsugai, "Entanglement entropy and the Berry phase in the solid state", Phys. Rev. B {\bf 73}, 245115 (2006).\\
11. H. T. Cui and J. Yi, "Geometric phase and quantum phase transition: Two-band model", Phys. Rev. A {\bf 78}, 022101 (2010).\\
12. T. Kitagawa, M.A. Broome, A. Fedrizzi, M.S. Rudner, E. Berg, I. Kassal, A. Aspuru-Guzik, E. Demler, and A.G. White, "Observation of topologically protected bound states in photonic quantum walks", Nature Commun. {\bf 3}, 882 (2012).\\
13. M. Verbin, O. Zilberberg, Y.E. Kraus, Y. Lahini, and Y. Silberberg, "Observation of Topological Phase Transitions in Photonic Quasicrystals", Phys. Rev. Lett. {\bf 110}, 076403 (2013).\\
14. M.C. Rechtsman, Y. Plotnik, J.M. Zeuner, A. Szameit, and M. Segev, "Topological creation and destruction of edge states in photonic graphene ", arXiv:1211.5683 (2012).\\
15. N. Goldman, J. Dalibard, A. Dauphin, F. Gerbier, M. Lewenstein, P. Zoller, and I. B. Spielman, "Direct imaging of topological edge states in cold-atom systems", PINAS  
{\bf 100}, 6736 (2013).\\
16. D.A. Abanin, T. Kitagawa, I. Bloch, and E. Demler, "Interferometric Approach to Measuring Band Topology in 2D Optical Lattices", Phys. Rev. Lett. {\bf 110}, 16530 (2013).\\
17. M. Atala, M. Aidelsburger, J.T. Barreiro, D. Abanin, T. Kitagawa, E. Demler, and I. Bloch, "Direct Measurement of the Zak phase in Topological Bloch Bands", arXiv:1212.0572v1\\
18. U. Peschel, T. Pertsch, and F. Lederer, "Optical Bloch oscillations in waveguide arrays", Opt. Lett. {\bf 23}, 1701(1998).\\ 
19. D. Christodoulides, F. Lederer, and Y. Silberberg, "Discretizing light behaviour in linear and nonlinear waveguide lattices",
Nature {\bf 424}, 817 (2003).\\
20. H. Trompeter, T. Pertsch, F. Lederer, D. Michaelis, U. Streppel,
A. Br\"{a}uer, and U. Peschel, "Visual observation of Zener tunneling", Phys. Rev. Lett. {\bf 96}, 023901 (2006).\\
21. A. Szameit and S. Nolte, "Discrete optics in femtosecond-laser-written photonic structures", J. Phys. B {\bf 43}, 163001 (2010).\\
22. I.L. Garanovich, S. Longhi,  A.A. Sukhorukov, and Y.S. Kivshar, "Light propagation and localization in modulated photonic lattices and waveguides", Phys. Rep. {\bf 518}, 1 (2012).\\
23. S. Longhi, "Classical and quantum interference in multiband optical Bloch oscillations", Phys. Rev. B {\bf 79}, 245108 (2009).\\
24. F. Dreisow, A. Szameit, M. Heinrich, T. Pertsch, S. Nolte, A. T\"{u}nnermann, and S. Longhi, "Bloch-Zener Oscillations in Binary Superlattices", Phys. Rev. Lett. {\bf 102}, 076802 (2009).\\
25. G. Corrielli, A.Crespi, G. Della Valle, S. Longhi, and O. Osellame, "Fractional Bloch oscillations in photonic lattices",
Nature Commun. {\bf 4}, 1555 (2013).\\
26.G.D. Marshall, A. Politi, J.C.F. Matthews, P. Dekker,
M. Ams, M.J. Withford, and J.L. O' Brien, "Laser written waveguide photonic quantum circuits", Optics Express {\bf 17}, 12546 (2009).\\
27. M.S. Rudner and L.S. Levitov, "Topological transition in a non-Hermitian
quantum walk", Phys. Rev. Lett. {\bf 102}, 065703 (2009).

\end{document}